\documentclass[fleqn,usenatbib]{mnras}

\usepackage{newtxtext,newtxmath}

\usepackage[T1]{fontenc}
\usepackage{bm}

\DeclareRobustCommand{\VAN}[3]{#2}
\let\VANthebibliography\thebibliography
\def\thebibliography{\DeclareRobustCommand{\VAN}[3]{##3}\VANthebibliography}

\usepackage{graphicx}	
\usepackage{amsmath}	

\newcommand{\concsection}[1]{\vspace{3mm}
\noindent \textbf{\textrm{{#1}}:}}

\title[Protohalos and Halo Assembly]{Protohalos and their connection to halo assembly, shape and structure}

\author[F. Nikakhtar et al.]{
Farnik Nikakhtar,$^{1}$\thanks{E-mail: farnik.nikakhtar@yale.edu}
Daisuke Nagai,$^{1}$
Marcello Musso,$^{2}$
Ravi K.~Sheth$^{3}$
\\
$^{1}$Department of Physics, Yale University, New Haven, CT 06511, USA\\
$^{2}$Departamento de Fisica Fundamental and IUFFyM, Universidad de Salamanca, Plaza de la Merced s/n, 37008 Salamanca, Spain\\
$^{3}$Center for Particle Cosmology, University of Pennsylvania, Philadelphia, PA 19104, USA
}

\date{Accepted XXX. Received YYY; in original form ZZZ}

\pubyear{\the\year{}}

\begin{document}
\label{firstpage}
\pagerange{\pageref{firstpage}--\pageref{lastpage}}
\maketitle

\begin{abstract}
Protohalos, primordial regions in the initial cosmic density field that evolve into dark matter halos, are crucial for understanding cosmic structure formation. Motivated by the potential to reconstruct protohalo positions and shapes from observed galaxies using a novel approach grounded in optimal transport theory, we revisit the relationship between the structural properties of protohalos and the assembly histories, concentrations, and final morphologies of their associated dark matter halos.
To better understand halo assembly, we introduce a new estimator defined by an integral over redshifts and compare its performance to $z_{50}$, the commonly used redshift at which half of the final halo mass is formed. We quantify protohalo structure using the three invariants of the inertia, deformation, and energy shear tensors. Although past research has correlated the first two invariants of the deformation and energy tensors with halo formation, our findings reveal that the third invariant also significantly correlates with halo assembly and final shape.
\end{abstract}




\section{Introduction}\label{intro}

The galaxies that we use to study galaxy formation and constrain cosmological models are embedded in dark matter halos.  This has motivated studies of how such halos assemble their mass.  In models where this assembly is driven by gravitational instability, halos are formed from the gravitational collapse of protohalo patches that were present in the initial fluctuation field.   
The conventional approach identifies protohalos with peaks in the spherically smoothed initial density fluctuation field \citep{bbks86, bm96, smt2001, psd13}.  However, recent work suggests that energy- rather than density-based characterizations may provide superior physical insight \citep{epeaks, mds2024}.  The energy shear tensor formalism uses both the magnitude and directional properties of collapse dynamics, and provides a principled way of going beyond the assumption of spherical protohalo shapes \cite{eshape23}.

In the current study, we are interested in quantifying how the structural properties of a protohalo correlate with the shape and assembly history of the final object.  This is because halo shape and structure are two of the leading contributors to the systematic error budget in cluster cosmology \citep{Allen2011,Blazek_shapebias2015,Bocquet2024}.  Therefore, in Section~\ref{sec:tensors}, we describe a number of tensors which we use to characterize the structure of protohalos.  Section~\ref{sec:sims} describes the simulations we use to illustrate our results and how we quantify the halo assembly and structure.  Section~\ref{sec:assembly} describes a new `integral' method for characterizing halo assembly, and compares it with previous `single-point' measures.  Section~\ref{sec:corrs} presents a number of correlations between protohalo properties and halo assembly history and shape.  Section~\ref{sec:energy} focuses on the protohalo parameters that play a key role in the energy peaks approach. Our main findings are summarized in Section~\ref{sec:conclusions}.

\section{Shape tensors and invariants} \label{sec:tensors}
First, we describe four tensors that quantify the structure of an object.  The mass and inertia tensors are the most familiar; to these, we add the deformation and energy tensors, which are believed to play a role in the dynamical evolution of the shape.  

\subsection{Mass and Inertia tensors}
The mass tensor, which quantifies the shape of an object, is defined by 
\begin{equation}
 M_{ij} \equiv \frac{1}{M}\sum_k m_k\, r_{k,i} r_{k,j},
 \label{eq:Mij}
\end{equation}
where $m_k$ is the mass of the $k$-th particle (in our simulations, all particles have the same mass), 
$r_k$ is its position vector relative to the center of mass, $r_{k,i}$ and $r_{k,j}$ are its components along the $i$-th and $j$-th axes, and $M\equiv \sum_k m_k$ is the total mass of the object.  
If $l_1^2\ge l_2^2\ge l_3^2$ are the ordered eigenvalues of the mass tensor, then a spherical shape has all $l_i=l_j$; an oblate object has $l_3\ll l_2$; and a prolate object has $l_1\gg l_2$.

The inertia tensor,  
\begin{equation}
 I_{ij} = \frac{1}{M}\sum_k m_k \left( r_k^2 \delta_{ij} - r_{k,i} r_{k,j} \right),
 \label{eq:Iij}
\end{equation}
where $\delta_{ij}$ is the Kronecker delta, highlighting the resistance to rotational motion around the center of mass. If $l_1^2, l_2^2$ and $l_3^2$ are the eigenvalues of the mass tensor, then the eigenvalues of the inertia tensor would be $l_2^2 + l_3^2$, $l_1^2 + l_3^2$ and $l_1^2 + l_2^2$.
Hence, if $l_1^2\ge l_2^2\ge l_3^2$ then $l_2^2 + l_3^2 \le l_1^2 + l_3^2 \le l_1^2 + l_2^2$. A convenient way to compare tensors for objects of different sizes is to normalize $I_{ij}$ (and similarly $M_{ij}$) by $3R^2/5$, the analytic value of the mass tensor for a uniform-density sphere of radius $R$. Alternatively, one may normalize by the total squared distance of all particles from the center of mass, $\sum_k r_k^2$, which removes the absolute size scale using the actual distribution and yields a dimensionless tensor that reflects only the relative shape of the system.

\subsection{Deformation tensor}

The deformation tensor \cite{doroshkevich1970}, defined as the spatial gradient of the displacement field, plays a key role in the `peak-patch' approach of \cite{bm96}. Since in the Zel'dovich approximation the displacement is given by the gradient of the potential perturbation $\phi$, the deformation tensor corresponds to the Hessian
\begin{equation}
 d_{ij} = \frac{\partial^2 \phi}{\partial x_i \partial x_j},
 \label{eq:Dij}
\end{equation}
where $x_i$ and $x_j$ are spatial coordinates. This tensor describes the tidal field, which arises from the second derivative of the potential and governs the stretching and compression experienced by an infinitesimal fluid element in different directions. The trace of this tensor corresponds to the Laplacian of the gravitational potential, which is directly related to the initial overdensity through the Poisson equation in the initial condition:
\begin{equation}
    \nabla^2\phi \equiv {\rm Tr}(d_{ij}) \propto \delta.
    \label{eq:pois}
\end{equation}
In the peak-patch approach, to capture the deformation of protohaloes, which are extended objects, one averages $d_{ij}$ over the fluid elements that contribute to the protohalo volume \citep{bm96, dts2013}:
\begin{equation}
    {\cal D}_{ij} = \frac{1}{N_G}\sum_k d_{ij}^{(k)}.
\end{equation}
The eigenvalues and eigenvectors of this smoothed deformation tensor characterize the nature of the tidal field, distinguishing between compressive and extensive forces along specific axes, with positive eigenvalues corresponding to compression and negative eigenvalues to expansion. This connection provides a direct link between the protohalo's shape and orientation and the large-scale structure of the universe, such as filaments, sheets, and voids, offering insights into how the environment influences the formation and evolution of cosmic structures.

\subsection{Energy tensor}

The energy overdensity tensor, or energy shear, plays a key role in the energy-peaks approach of \cite{epeaks}.  For a generic comoving volume $V$ it is defined by: 
\begin{equation}
    \epsilon_{ij} \equiv 3\frac{ \int_V \mathrm{d}^3r \, \rho(\mathbf{r}, t) \left(r_i - r_{\mathrm{cm}i}\right) \left(\nabla_j \phi - [\nabla_j \phi]_{\mathrm{cm}}\right)}{\int_V \mathrm{d}^3r \, \rho(\mathbf{r}, t) \,|\mathbf{r} - \mathbf{r}_{\mathrm{cm}}|^2},
    \label{eq:Eij}
\end{equation}
\citep{epeaks, mds2024}. Here, $\mathbf{r}_{\mathrm{cm}}$ is the center of mass position (i.e., it is the trace of equation~\ref{eq:Mij}) and $\phi$ is the same potential that is related to the density contrast $\delta$ by the Poisson equation (equation~\ref{eq:pois}), so $[\nabla \phi]_{\mathrm{cm}}$ corresponds to the gravitational acceleration at the center of mass. 

The tensor $\epsilon_{ij}$ encodes the coupling between the density and velocity fields, characterizing the anisotropic energy distribution and providing a framework for analyzing the dynamical response of protohalos to their surrounding environment.  In the initial conditions, $\nabla\phi = \bm{v}/fDH$, where $f\equiv d\ln D/d\ln a$ and $D$ is the linear theory growth factor.  Hence, for each protohalo, we estimate 
\begin{equation}
    \epsilon_{ij} \equiv -\frac{3}{fDH} \,
    \frac{\sum_k r_{k,i} \,v_{k,j}}{\sum_k r_k^2},
    \label{eq:estEij}
\end{equation}
where $r_{k,i}$ and $v_{k,j}$ are the components of the $k$th particles position and velocity with respect to the center of mass, and the sum is over all its particles \citep{rotinv}.

Since $\epsilon_{ij}$  depends on one derivative of the potential rather than two, it is slightly more robust to measure than ${\cal D}_{ij}$.  In addition, for protohalos, all three eigenvalues of $\epsilon_{ij}$ are positive, whereas this is not the case for ${\cal D}_{ij}$ \citep{mds2024}.

\subsection{Invariants of tensors}
A $3\times 3$ tensor has three eigenvalues, and it is conventional to combine them into three invariants.  For $\epsilon_{ij}$, these are the trace, often called the `(energy) overdensity', the amplitude of the traceless shear, and the traceless determinant:  
\begin{equation}
\begin{split}
    \epsilon &\equiv \sum_i \lambda_i, \\
    q^2 & \equiv \frac{3}{2} \sum_i\left(\lambda_i-\epsilon / 3\right)^2, \\
    U^3 & \equiv \sum_i\left(\lambda_i-\epsilon / 3\right)^3=3 \prod_i\left(\lambda_i-\epsilon / 3\right) .
    \label{eq:invariants}
\end{split} 
\end{equation}
In a Gaussian random field, the trace is Gaussian distributed, the (amplitude of the) traceless shear is chi-squared with 5 degrees of freedom, and the combination $(9/2)\, (U^3/q^3)$ is uniform.  
Whereas $q^2$ and $U^3$ are nonlinear combinations of the $\lambda_i$, we will also be interested in the following linear combinations:   
\begin{align}
    e\epsilon &\equiv \frac{\lambda_1-\lambda_3}{2},\qquad 
    p\epsilon \equiv \frac{\lambda_1+\lambda_3-2\lambda_2}{2}, \quad{\rm and}\\
    v_\pm &\equiv (\lambda_1-\lambda_3) \pm (\lambda_2 - \lambda_3),
    \label{eq:epv}
\end{align}
where $\lambda_1\ge\lambda_2\ge\lambda_3$.  The variables $e$ and $p$ (with $-e\le p\le e$) are defined by differences with respect to the middle eigenvalue; instead, $v_\pm$ uses differences with respect to the smallest.  The former were used by the peak-patch approach of \cite{bm96}, where the deformation tensor is the fundamental quantity; the latter are more natural in the energy peaks approach of \cite{rotinv}, in which the physics is driven by $\epsilon_{ij}$, particularly because, for protohalos, all three eigenvalues of $\epsilon_{ij}$ are positive, whereas this is not the case for $D_{ij}$ \cite{mds2024}.
Finally, notice that the trace of the inertia tensor is twice that of the mass tensor, and, except for this factor of two, $e$ and $p$ for the mass tensor and the inertia tensor are equal.  Similarly, $q^2$ and $|U^3|$ for the two tensors are equal.

\section{Protohalos in simulations}\label{sec:sims}

\begin{figure}
    \centering
    \includegraphics[width=\linewidth]{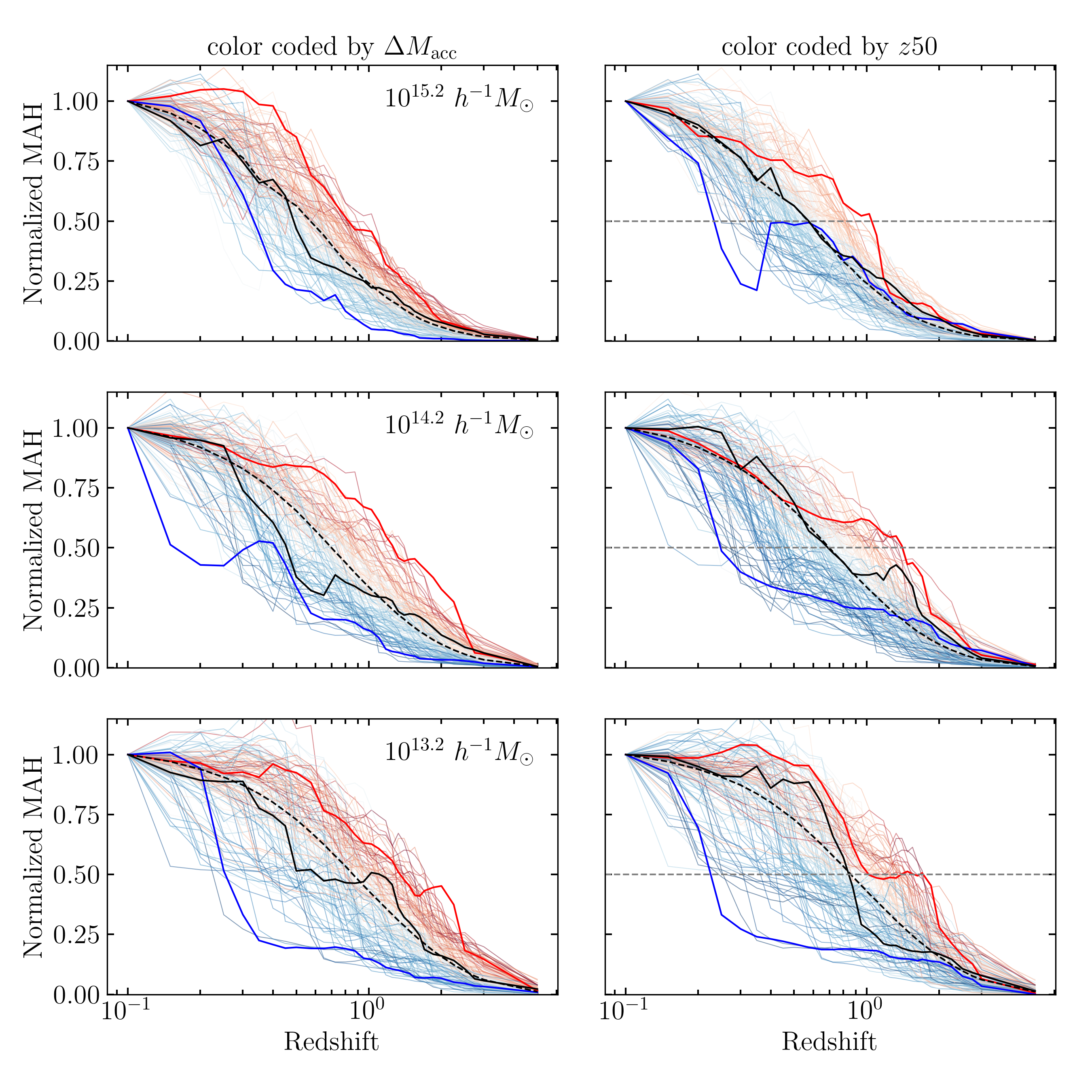} \\
    \caption{Mass accretion histories color coded by $\Delta M_{\rm acc}\equiv A$ of equation~(\ref{eq:MAH}) (left) and $z_{50}$ (right), for three narrow bins in halo mass (top to bottom). Dashed curve, same in each pair of panels, shows the median mass at each $z$.  Solid black curve shows the accretion history of the object with the median $A$ and $z_{50}$; solid red and blue curves show the histories of the objects with the most extreme values.}
    \label{fig:MAH}
\end{figure}

We conduct our analysis using a single simulation box from the {\sc{AbacusSummit}} suite \cite{Garrison2021, Maksimova2021}, with a volume of \((2 \, \mathrm{Gpc}/h)^3\), simulated under a fiducial flat $\Lambda$CDM cosmology with parameters $(\Omega_m, \Omega_b, h) = (0.3152, 0.0507, 0.6736)$ and $2.1 \times 10^9 \, \mathrm{M}_\odot/h$ particles. Halos are identified at $z = 0.1$ using the {\texttt{CompaSO}} halo finder \cite{Hadzhiyska2022}, and the particle IDs are matched to their positions in the initial conditions at $z^{\textrm{IC}} = 99$. The Lagrangian regions occupied by these particles define the protohalos.

For each protohalo, we have access to the position and velocity of a subsample of its particles on a grid in configuration space. Additionally, we have the initial density field sampled on a lower-resolution \(2048^3\) grid. To compute the inertia, mass, and energy tensors, we use the configuration-space positions and velocities of these particles. However, for the deformation tensor, which requires the Hessian of the gravitational potential, we do not estimate derivatives directly in configuration space. Instead, we first perform a Fast Fourier Transform (FFT) to obtain the density contrast $\delta(k)$ on a Fourier-space grid. We then compute the second derivatives of the potential by evaluating $k_i k_j\,\delta(k)/k^2$ at each Fourier-space grid point. Finally, we apply an inverse FFT to transform the result back to real space and average over all grid cells contained within the Lagrangian volume of the protohalo.

To construct the merger tree, we used the methodology outlined in \cite{Bose2022}, which enables high-fidelity reconstruction of halo mass accretion histories and allows for the identification of significant evolutionary classes of halos, such as smooth accretion vs. merger-dominated halos or early-collapse halos vs. late-forming halos.

\begin{figure}
    \centering
    \includegraphics[width=\linewidth]{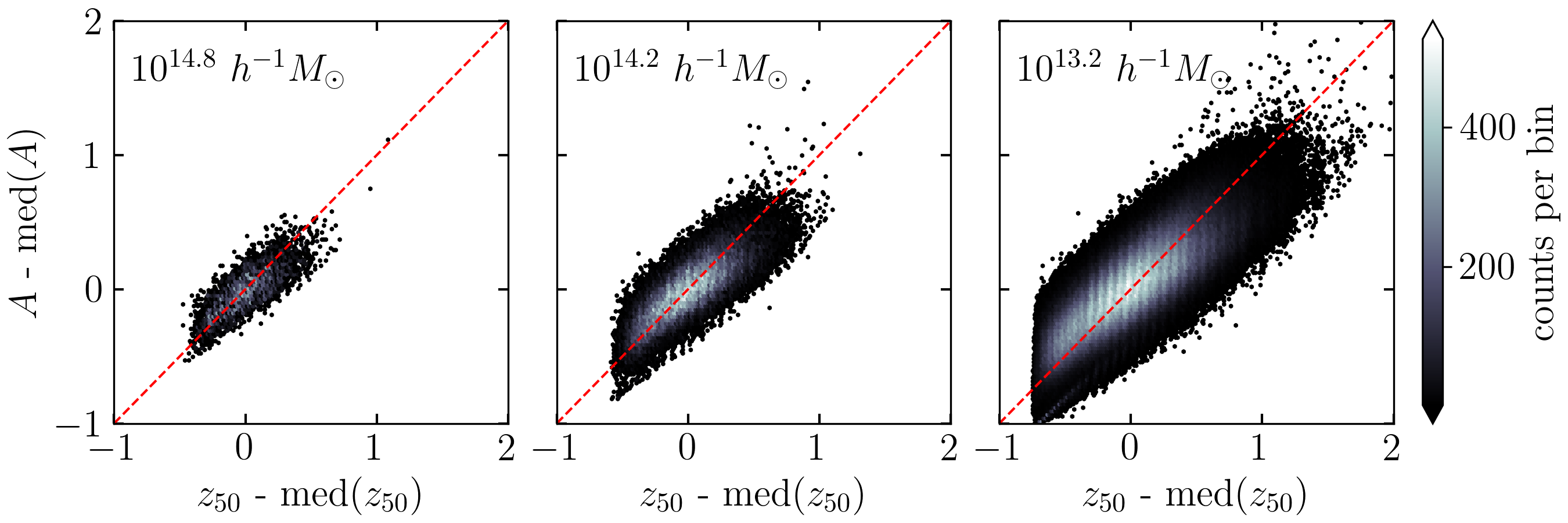}
    \caption{Difference from median history as quantified by $A$ is well correlated with that for $z_{50}$.}
    \label{fig:W1-z50}
\end{figure}

\section{Single-time and integrated measures of assembly history}\label{sec:assembly}

\begin{figure*}
  \includegraphics[width=\textwidth]{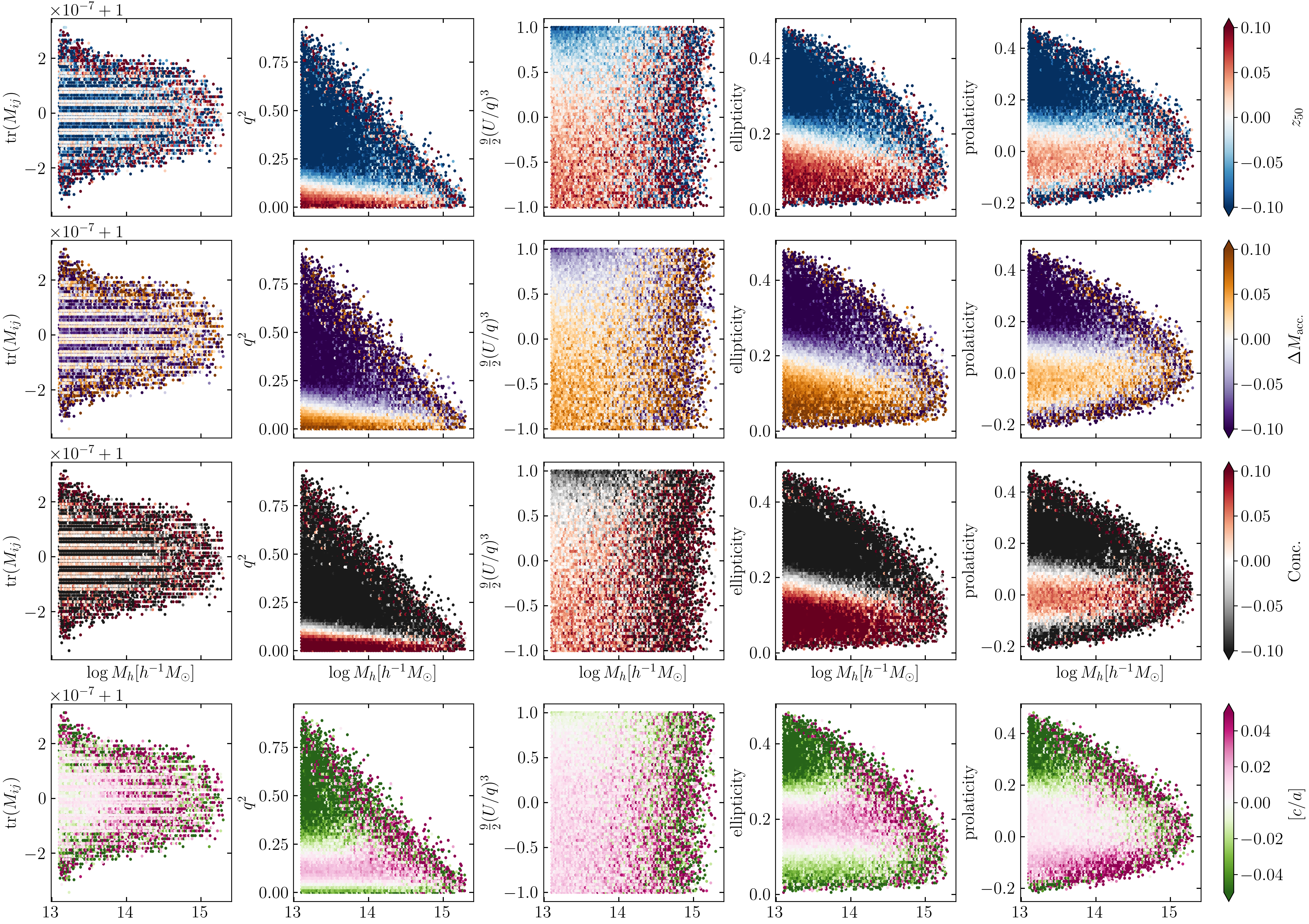}
  \caption{Mean halo property (top to bottom shows $z_{50}$, our new integrated measure of assembly history $A$, final concentration and axis ratio $[c/a]$) as a function of halo mass and invariants of the protohalo mass tensor (left to right shows trace, traceless shear, dimensionless ratio of traceless determinant and shear, trace times ellipticity and trace times prolateness).  In all cases, the median trend with mass has been removed, so the white band (zero residual) shows the median trend, and the other colors show how the residual from the median relation depends on invariant.} 
  \label{fig:tensorM}
\end{figure*}

Halo assembly is a stochastic process, so it is useful to have a simple way of characterizing or quantifying assembly history.  The redshift $z_{50}$, at which half of the final mass was first assembled, is a popular choice, especially because it has been shown to correlate with the central concentration of the final object \cite{nfw1997}.  However, because it corresponds to a single time, estimating $z_{50}$ requires reasonably closely spaced outputs.  To overcome these limitations, we instead use an \emph{integrated deviation} of the mass accretion history (MAH) from the median MAH at each redshift. Specifically, we define the normalized mass accretion history for a given halo as $m(z) = M(z)/M(z_{\rm{fin}})$, where $M(z)$ is the halo mass at redshift $z$, and $M(z_{\rm{fin}})$ is its final mass (typically at $z = 0$). We then compare this curve to the median normalized accretion history $\tilde{m}(z)$, computed from a population of halos with the same final mass. The integrated difference is given by

\begin{equation}
    A = \int_{z_{\rm init}}^{z_{\rm fin}} dz\, \left[ m(z) - \tilde{m}(z) \right].
    \label{eq:MAH}
\end{equation}

This quantity $A$ reflects the relative assembly history of a halo: If $A>0$, the halo assembled its mass earlier than average; if $A<0$, the halo assembled later than average. Unlike $z_{50}$, this metric incorporates the entire growth history and is less sensitive to the temporal resolution of the output.

\begin{figure*}
  \includegraphics[width=\textwidth]{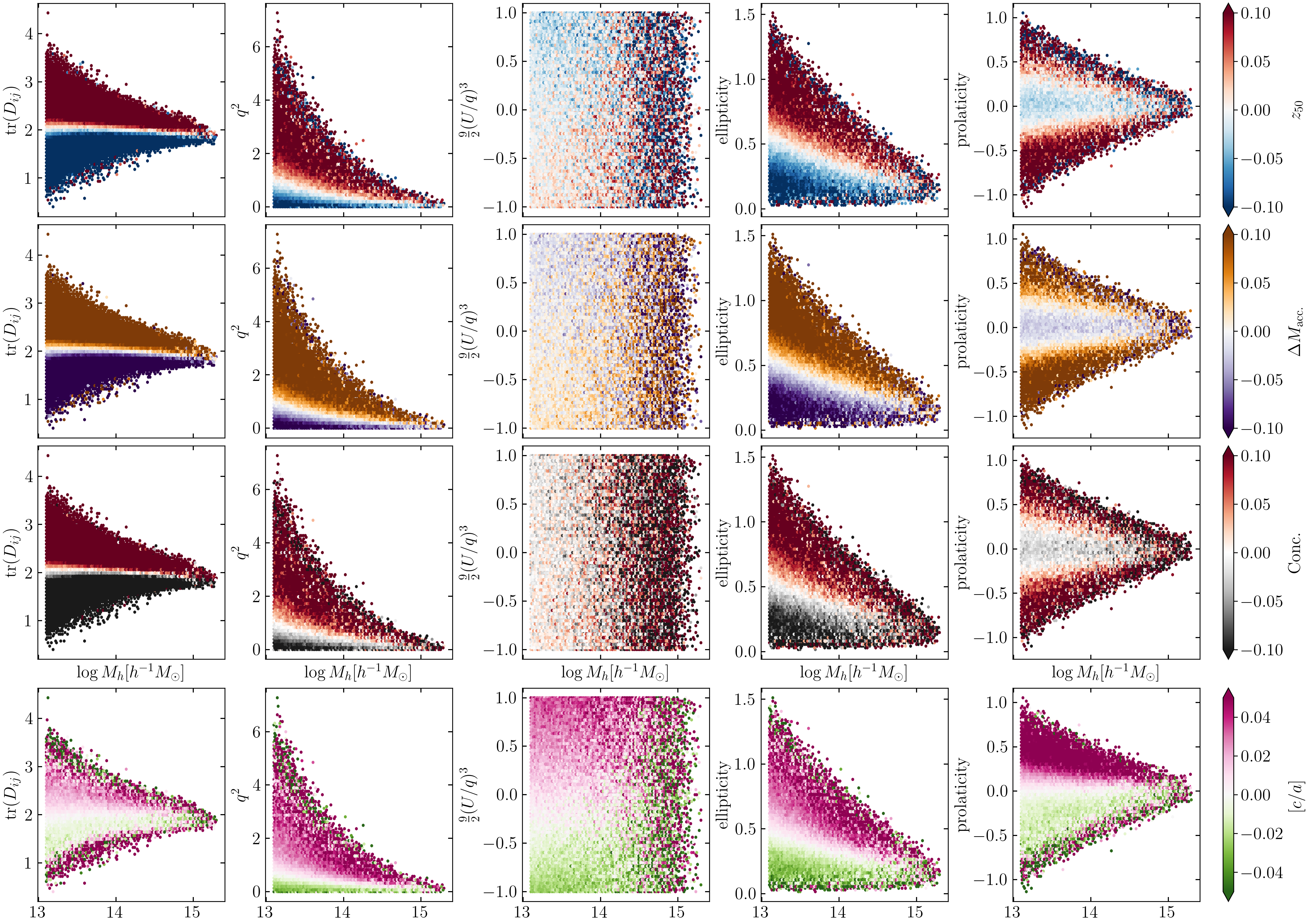}
  \caption{Same as previous figure, but now for the invariants of the protohalo deformation tensor.}
  \label{fig:tensorD}
\end{figure*}

\begin{figure*}
  \includegraphics[width=\textwidth]{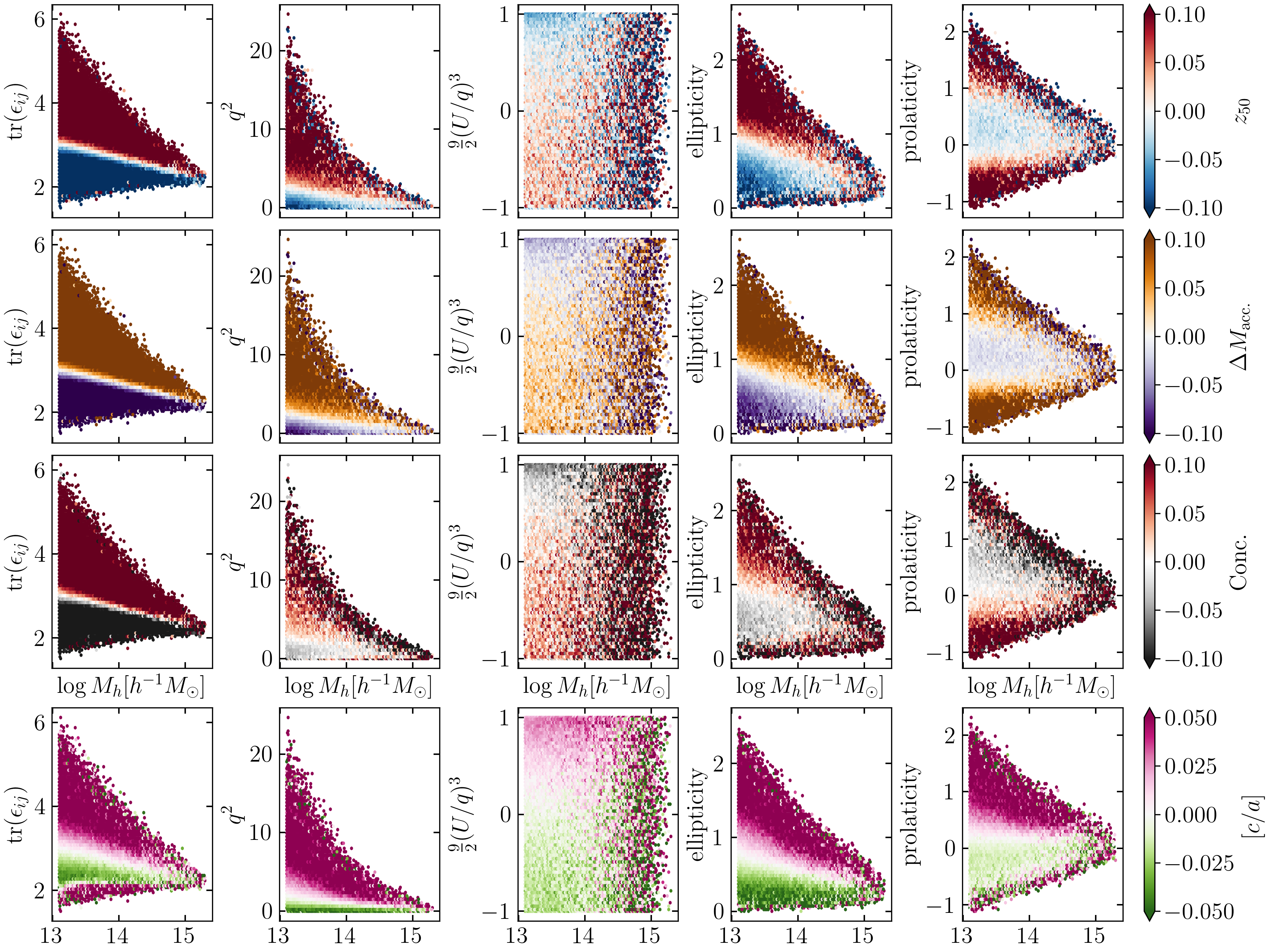}
  \caption{Same as previous figure, but now for the invariants of the protohalo energy shear tensor.}
  \label{fig:tensorE}
\end{figure*}
Figure~\ref{fig:MAH} shows a random subset of halo accretion histories in three narrow mass bins.  The mass growth tracks have been colored by their values of $A$ (left-hand panels) and $z_{50}$ (right-hand panels).  The dashed black curve (the same in both panels for a given mass bin) traces the median value of $M(z)/M_0$ at each $z$.  At higher masses (top panels) this curve crosses 0.5 at lower redshifts, illustrating the tendency for more massive halos to assemble later.  Over this range of masses, the median $z_{50}\approx 0.725 - 0.125\,\log_{10}(M/10^{14}h^{-1}M_\odot)$.  The solid black curves show the accretion history for the halo with the median $A$ (left) and $z_{50}$ (right) for that same mass bin, with red and blue showing the histories of objects that had earlier or later assembly than the median $A$ or $z_{50}$, and the thick red and blue curves showing the histories associated with the most extreme values.  It is reassuring that, in each panel, the solid black curve matches the dashed one reasonably well: $A$ is as reasonable a choice as $z_{50}$ for parameterizing the entire history.  In fact, Figure~\ref{fig:W1-z50} shows that these two measures of halo assembly are well correlated.  In what follows, we will study how $A$ and $z_{50}$ correlate with other protohalo properties.

\section{Correlation of halo assembly history and shape with protohalo properties} \label{sec:corrs}

We begin with a study of how the halo properties are correlated with the mass tensor ${\cal M}$ of the protohalo.  Each panel in Figure~\ref{fig:tensorM} shows the distribution of an invariant (trace, traceless shear $q^2$, $U^3/q^3$, ellipticity, and prolateness) versus halo mass (the plot uses trace times $e$ and $p$, the quantities on the right hand side of equation~\ref{eq:epv}, rather than $e$ and $p$ themselves).  In the top set of panels, the coloring shows the residual, $z_{50}$ minus the median $z_{50}$, for its mass bin, with white corresponding to (no difference from) the median value.  
The other panels are colored by the corresponding residuals in $A$, concentration and axis ratio (short-to-long).  (For reference, over the mass range shown, the median values scale as $\log_{10}({\rm conc}/3.66) \approx -0.042\, \log_{10} (M/10^{14}h^{-1}M_\odot)\approx [c/a]_{\rm Eul}-0.57$.)  The protohalo patches are nearly spherical, so there is little information on the trace; we have kept this left-most panel for ease of comparison with Figures~\ref{fig:tensorD} and~\ref{fig:tensorE}, showing the corresponding results for the deformation and energy shear tensors, ${\cal D}$ and ${\cal E}$.

Before commenting on the color coding, note that all quantities except $U^3/q^3$ cover a wider range of values at lower masses.  Earlier work \cite{dts2013} had shown that, for the deformation tensor, the range scales approximately as $\sigma(M)$ for the trace and $\sigma^2$ for the traceless shear.  Evidently, this is true for the other tensors as well.  

We noted earlier that white shows the median trend with mass.  Clearly, at smaller masses, the median $z_{50}$, $A$ and concentration, which are known to be correlated with one another, occur at slightly larger $q$ and $e$, approximately consistent with the $\sigma$-scaling we mentioned earlier.  On the other hand, the median $c/a$ clearly occurs at two values of $q$ and $e$ (the panels show two white bands rather than a single one). The roundest objects have $q^2\sim 0.1$ and $e\sim 0.2$, with $c/a$ decreasing in the larger and smaller $q$ or $e$.  In other words, the roundest objects today did {\em not} form from the roundest initial patches.  This is consistent with recent work suggesting that the final shape is a consequence of the initial shape and the asymmetry in the initial amounts by which it was squeezed \cite{eshape23}.  Moreover, at low masses, the objects that end up being roundest tend to be the objects with median formation times.  

Likewise, objects that form the earliest and are the most concentrated tend to have $p\sim 0$, with later formation associated with larger and smaller $p$.  Note that $p=0$ has trace equal to $3\lambda_2$, or $\lambda_1-\lambda_2 = \lambda_2-\lambda_3$:  i.e., objects with two equal eigenvalues are {\em not} the earliest to form.  

The white bands in Figures~\ref{fig:tensorD} and~\ref{fig:tensorE} show similar trends, except that now it is the latest forming objects that have $p_{\cal D}=0$ (for the deformation tensor) and $p_{\cal E}\sim 0.1$ (for the energy shear tensor).  This is part of a broader theme: many of the trends are similar for $\cal{D}$ and $\cal{E}$, but they have the opposite sign for $\cal{M}$.  

Before we consider the other panels, consider the trace.  For $\cal{D}$, the distribution of $\delta$ values is symmetric about a mean value, and the median $z_{50}$, $A$ and the concentration all occur at this same $\delta$ for all $M$.  For $\cal{E}$, the distribution of $\epsilon$ is asymmetric (there is an effective lower limit), consistent with \cite{mds2024}, and the median formation time occurs at higher $\epsilon$ values as $M$ decreases.  Likewise, the distribution of prolateness $p_{\cal E}$ is clearly asymmetric, in contrast to $p_{\cal{D}}$ which is much more symmetric around $0$.

\begin{figure}
    \centering
    \includegraphics[width=0.95\linewidth]{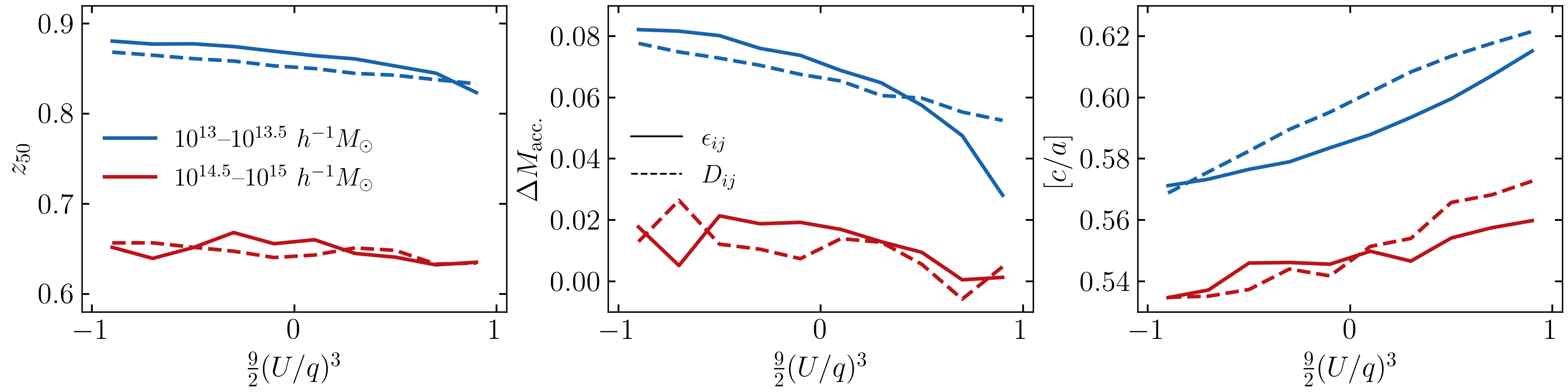}
    \caption{Correlation between assembly history (left, middle) and shape (right) with the third invariant of the energy (solid) and deformation (dashed) tensors, for two bins in protohalo mass.  (For this plot only, we have not removed the mean trend:  massive halos formed later and are less round.)  Larger values of the invariant are associated with later formation and rounder shapes, especially at lower masses. }
    \label{fig:3rd}
\end{figure}

Turning now to the color coding, Figure~\ref{fig:tensorM} shows that protohalos with larger $q^2_{\cal M}$, $e_{\cal M}$ and $p_{\cal M}$ are associated with halos that formed later (have smaller $z_{50}$ and $A$), are less concentrated and are more aspherical (smaller $c/a$).  In addition, there is a hint that these trends are also true for lower-mass objects with larger $U_{\cal M}^3/q_{\cal M}^3$ (i.e., the color gradient is similar to the other panels).  In contrast, for the deformation and energy shear tensors, objects with larger trace, $q^2$ and $e$ formed earlier, and are more concentrated and rounder (i.e., these trends are opposite those for the inertia tensor).  In addition, while there is little correlation between $U^3/q^3$ and halo properties at high masses, the lower mass objects with larger $U^3/q^3$ formed later and are less concentrated but are rounder (i.e., the color gradient is inverted with respect to the other panels, so that it is more similar to that for the mass tensor).  This is highlighted in Figure~\ref{fig:3rd}.
The correlation between this third invariant and the final shape is particularly striking; we discuss it in more detail in the next section.  

\section{On the energy peaks approach}
\label{sec:energy}
The previous section showed how the halo shape and assembly correlate with protohalo properties and how these correlations depend on mass.  Not all the correlations shown there are independent: some may be consequences of other more fundamental correlations.  

The trend for larger $\delta$ to form earlier and be more concentrated is qualitatively consistent with previous work showing that larger $\delta$ has larger $z_{50}$ \cite{borzy_colltime14} and models which suggest that this is also true for $\epsilon$ \cite{rotinv}.  Physically, this is because objects that assembled their mass earlier, when the universe was denser, tend to have more centrally concentrated profiles even at $z=0$ \cite{nfw1997}, so the protohalo correlations in Figures~\ref{fig:tensorM}--~\ref{fig:tensorE} provide an explicit link between the properties of a protohalo and the structure of the corresponding virialized halo at later times.

In halo formation models, large $\delta$ or $\epsilon$ are associated with larger shear \cite{smt2001, epeaks}.  Figures~\ref{fig:tensorD} and~\ref{fig:tensorE} show that, in fact, objects with the highest shear tend to have the largest $z_{50}$.  So, is $z_{50}$ determined by $q_{\cal D}$ or $q_{\cal E}$?  We now argue that the trends with shear and the formation history are distinct.

In generic positions of a Gaussian field, the trace and traceless shear are independent.  However, they are strongly correlated in protohalo patches: protohalos with large $\delta$ tend to have large $q_{\cal D}$, and large $\epsilon$ tend to have large $q_{\cal E}$.  For example, in the energy peaks approach, the trace of ${\cal E}$ is well approximated by 
\begin{equation}
    \epsilon \approx \sqrt{\epsilon_c^2 + v_+^2},
    \label{eq:epsv+}
\end{equation}
with $\epsilon_c\approx 2.2$ \cite{rotinv}; there is very little scatter around this relation.  So, it is reasonable to ask whether the small scatter that remains correlates with halo assembly or shape (or both).  

\begin{figure}
    \centering
    \includegraphics[width=0.95\linewidth]{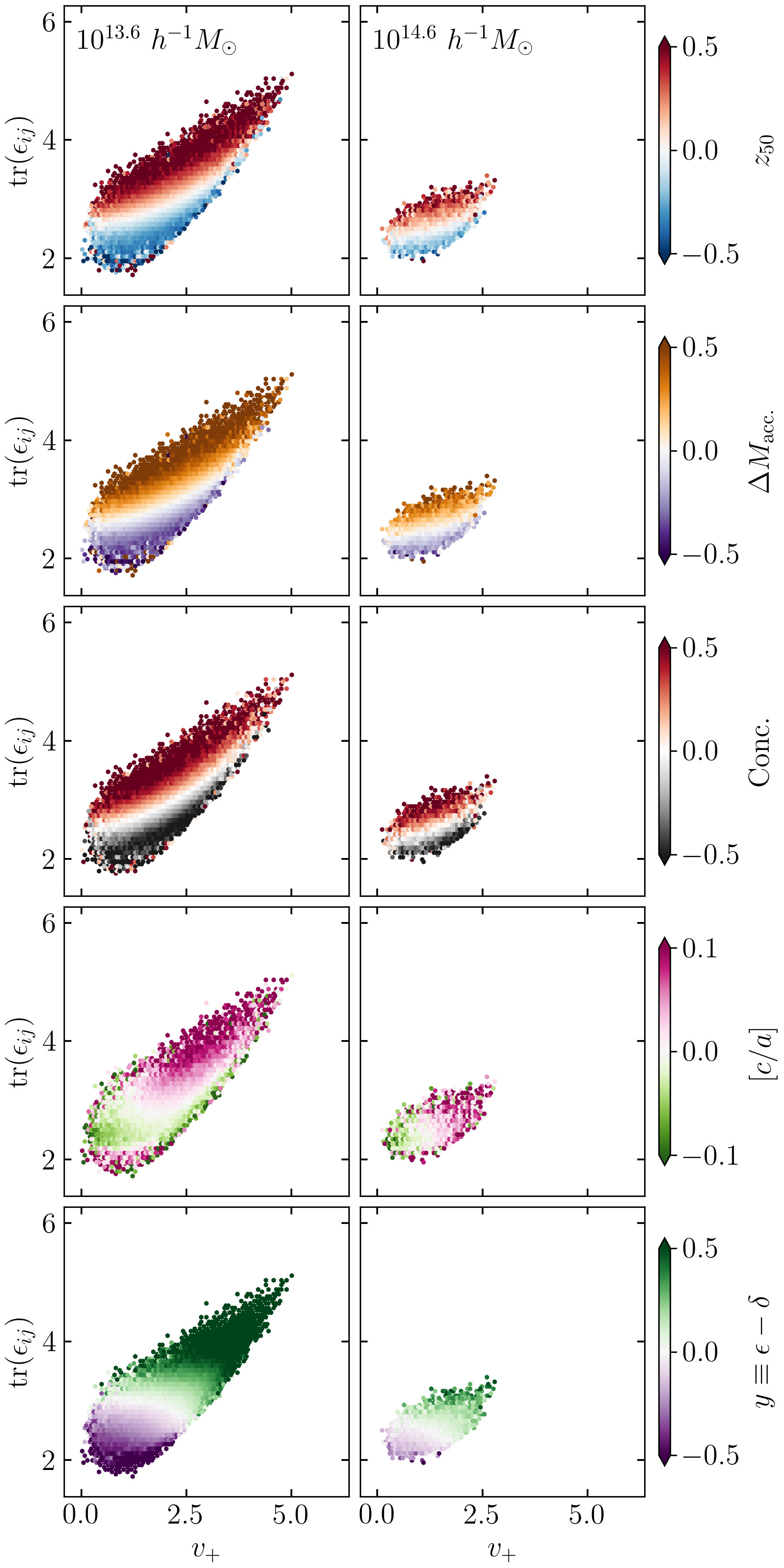}
    \caption{Dependence of $z_{50}$ and $A$ (top two panels), concentration (third from top), shape (second from bottom) and slope of the energy profile $y\equiv \epsilon - \delta$ (bottom) on $\epsilon$ and $v_+$.   }
    \label{fig:barrierczy}
\end{figure}

To illustrate, Figure~\ref{fig:barrierczy} shows $\epsilon$ versus $v_+$ for a sample of halos of low and high mass (left and right panels).  First, note that, especially at lower masses, the full range in the $\epsilon$ values is large compared to the range in the fixed $v_+$.  (The slight curvature of the $\epsilon-v_+$ correlation is captured by the expression above.)  This justifies the assertion that $\epsilon$ and $v_+$ are quite strongly correlated.  

In the top two panels, we have weighted objects by $z_{50}$ and $A$.  Evidently, the scatter around the mean $\epsilon-v_+$ relation correlates with the assembly history: protohalos with $\epsilon > \sqrt{\epsilon_c^2 + v_+^2}$ tend to have larger $z_{50}$.  The next (third) panel shows that the final concentration behaves similarly, consistent with the known $z_{50}-$concentration correlation.  The fourth panel shows that the shape does not depend strongly on the distance from the mean $\epsilon-v_+$ correlation, although, for more massive halos, there appears to be a tendency for protohalos with larger $v_+$ to become rounder halos.  This appears to be qualitatively different from the lower mass sample, just as the correlation between halo shape (and assembly) with the third invariant $U^3/q^3$ appears at smaller masses only.  These qualitative differences between low- and higher-mass objects are suggestive: in the energy peaks approach, the protohalo structure changes from being dominated by the requirement that all eigenvalues of ${\cal E}$ be positive (at lower masses) to their sum (the trace) being large enough to ensure collapse at present time (at higher masses) \cite{mds2024, rotinv}.   

\begin{figure*}
    \centering
    \includegraphics[width=0.95\linewidth]{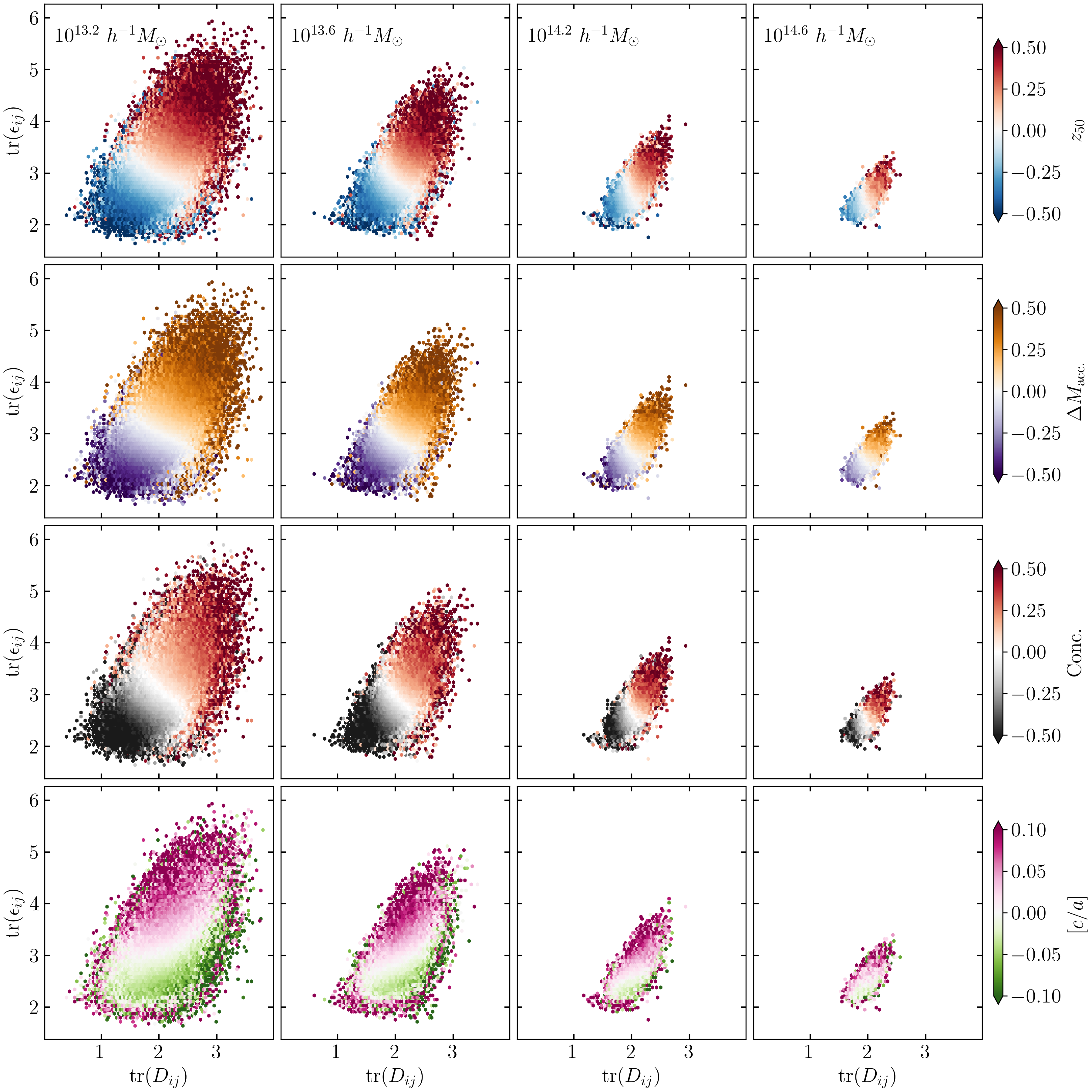}
    \caption{Halo assembly history (top two panels), concentration (third from top) and shape (bottom), as a function of location in the $\epsilon-\delta$ plane, for a number of bins in halo mass.  }
    \label{fig:ed}
\end{figure*}

Finally, in the bottom panel, the objects have been weighted by $y\propto \epsilon - \delta$ (as for the panels above it, we have subtracted the mean for the mass bin).  This is potentially interesting because the positivity constraint makes $v_+$ correlate with $\epsilon$, but does not imply anything for $y$. On the other hand, $y$ and $\epsilon$ are always correlated.  Here, the figure shows that color gradients are horizontal rather than perpendicular to the mean relation, suggesting that $y$ is determined by $\epsilon$ rather than $v_+$.  

Along these lines, Figure~\ref{fig:ed} shows how halo assembly history, concentration and shape correlate with location in the $\epsilon-\delta$ plane.  Lines of fixed $y\propto\epsilon-\delta$ would have positive slope, different from the median $z_{50}$ trends, but more similar to the trends with [$c/a$].  Moreover, the plot shows that the assembly history (and concentration) change from being determined by $\delta$ (colors are approximately vertical stripes) at large masses to $\epsilon$ (colors are approximately horizontal stripes) at smaller masses.  This is consistent with the point we made above about there being two regimes:  one dominated by the positivity constraint, and the other by a constraint on the trace.

\begin{figure}
    \centering
    \includegraphics[width=0.475\linewidth]{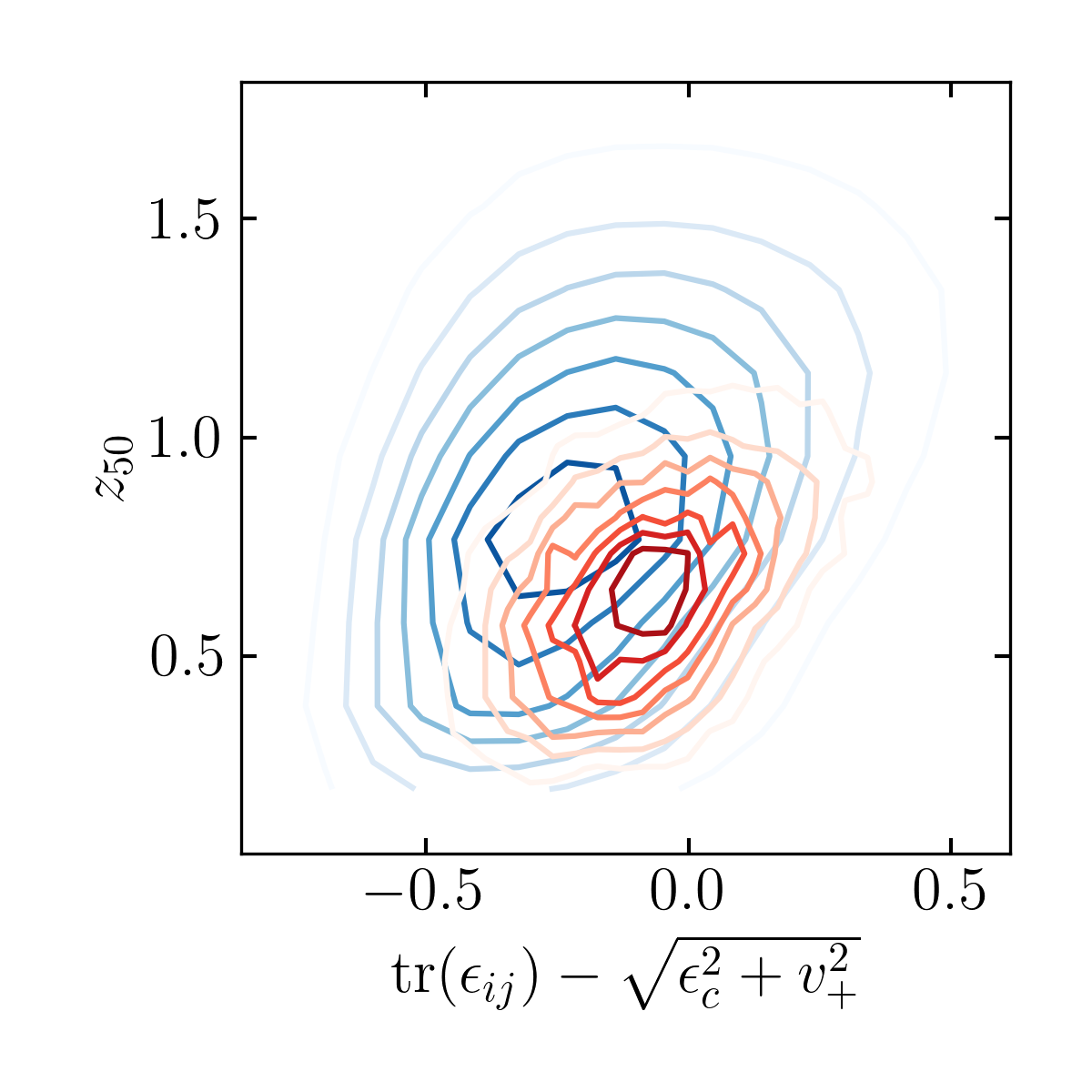}
    \includegraphics[width=0.475\linewidth]{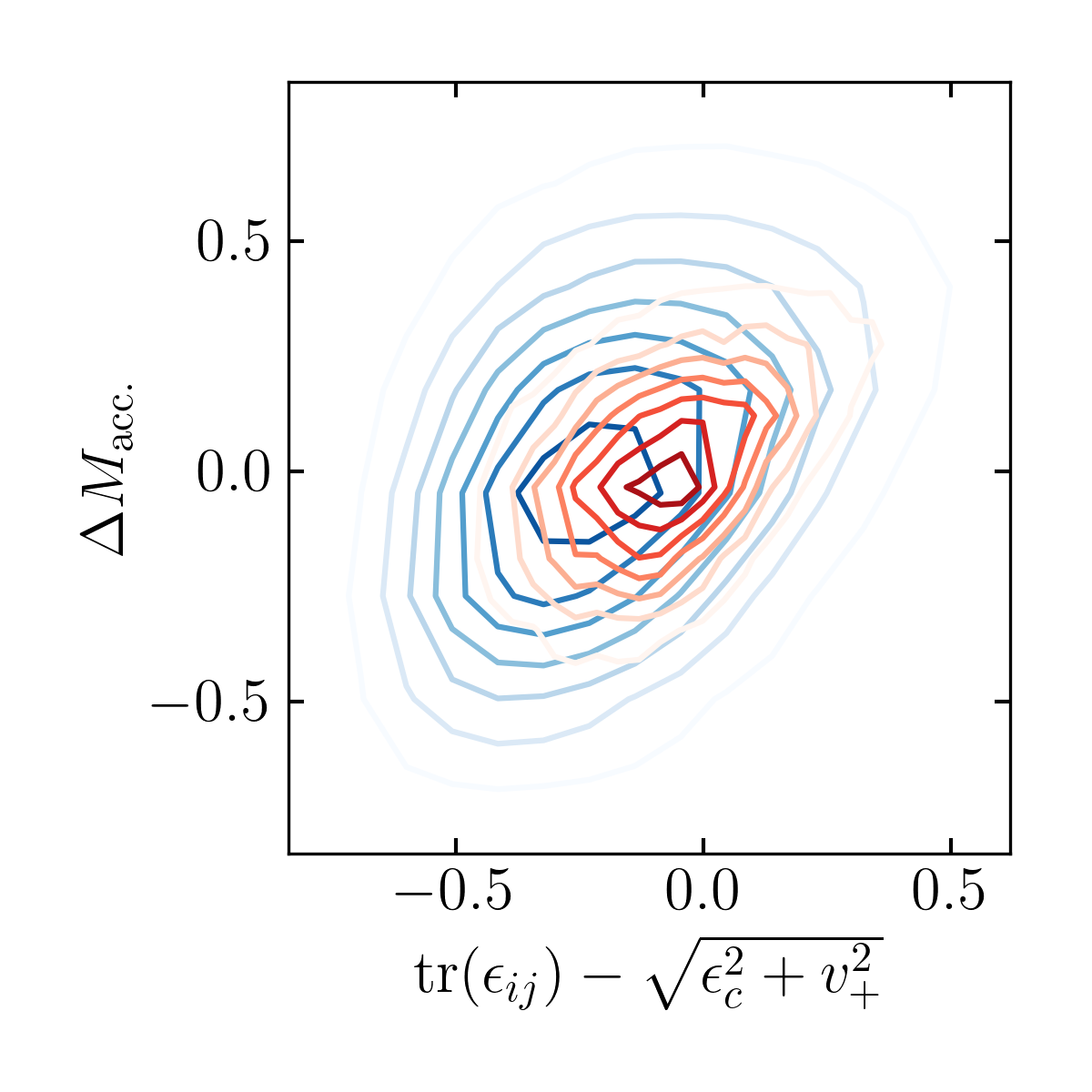}
    \includegraphics[width=0.475\linewidth]{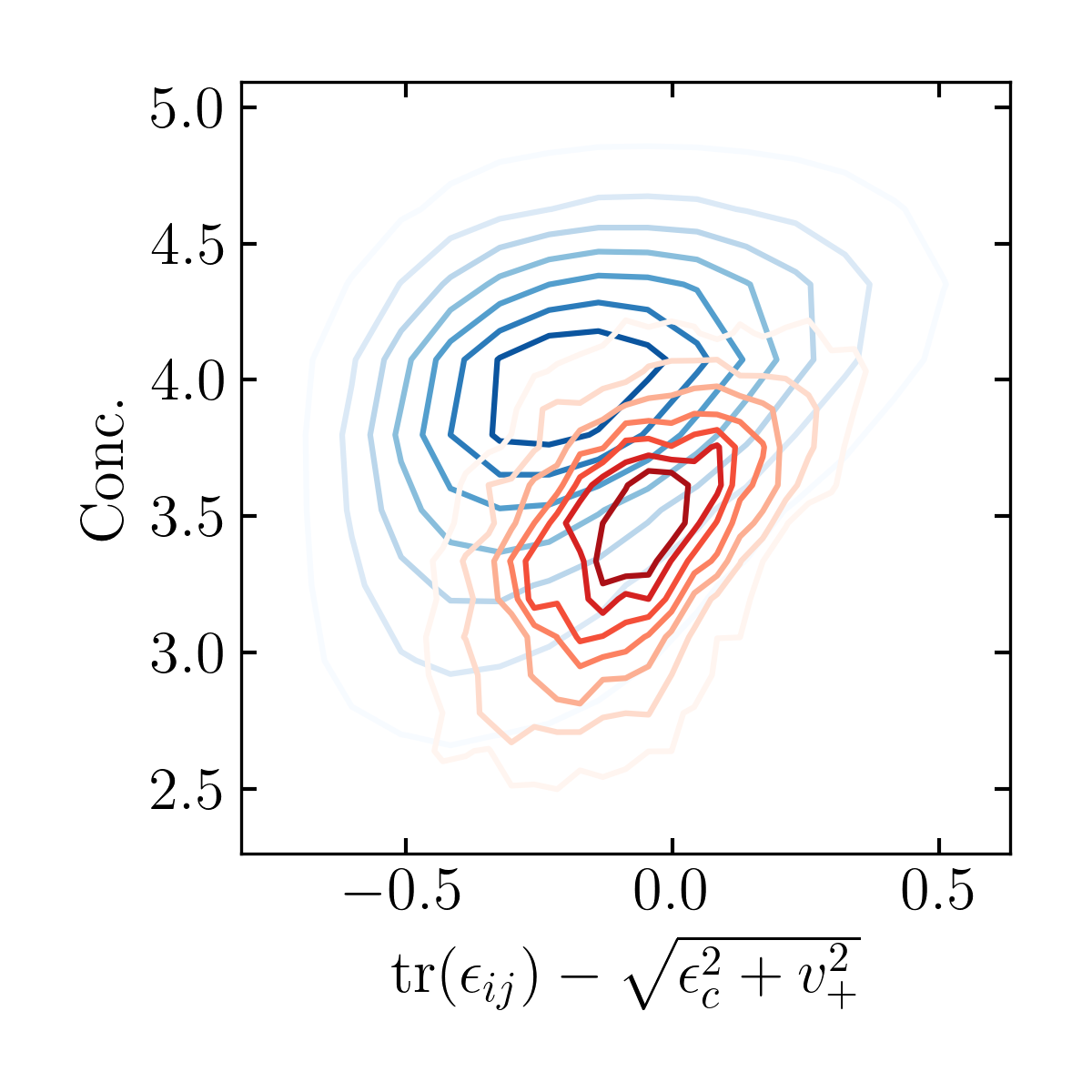}
    \includegraphics[width=0.475\linewidth]{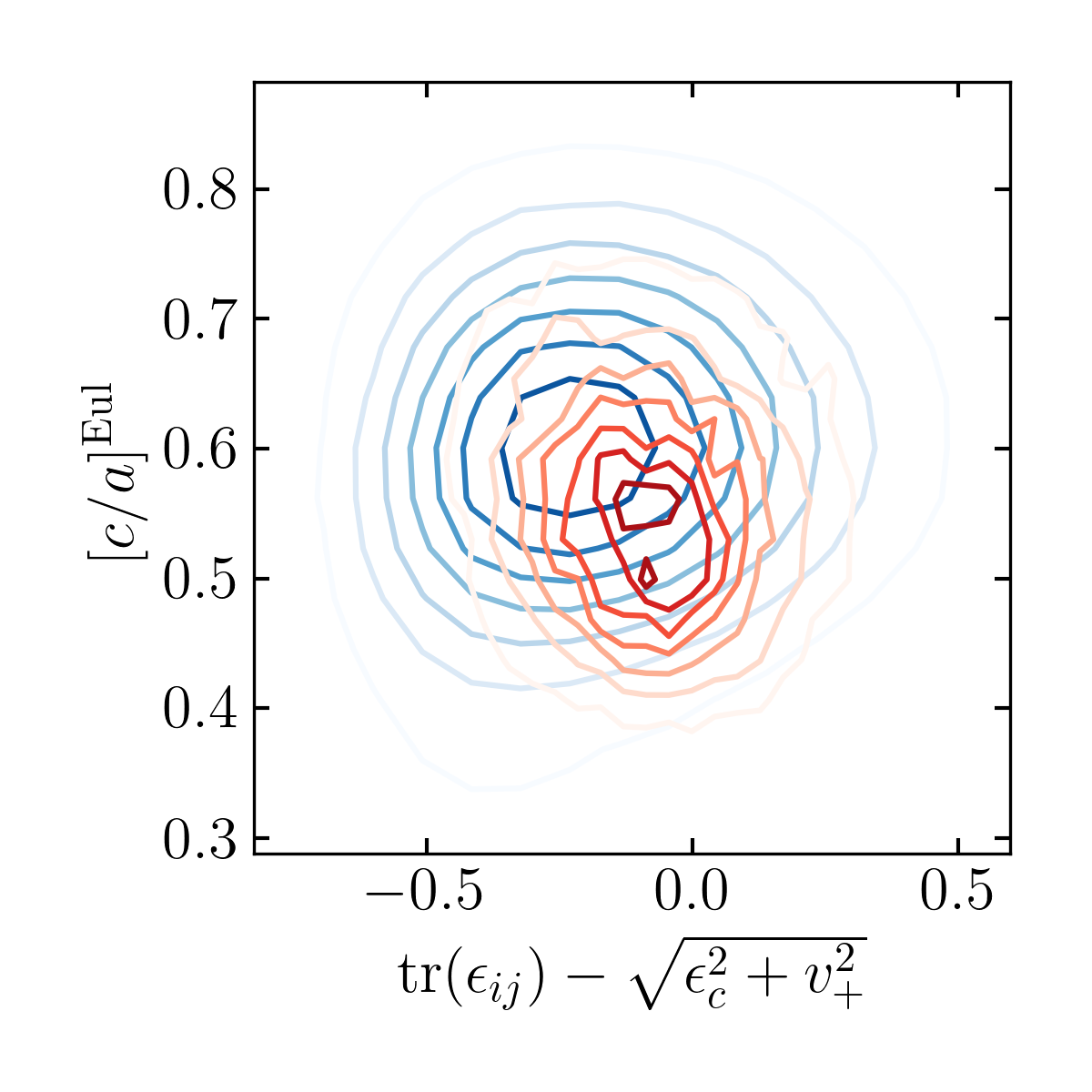}
    \caption{At fixed mass (blue and red contours show halos with masses in a narrow range around $\log_{10}(M/h^{-1}M_\odot) = 13.6$ and 14.6), residuals from the mean threshold correlate with formation time (top):  protohalos with larger initial over-densities tend to have larger $z_{50}$.  This leaves a signature in the concentration of the evolved halo (bottom left) but less so for its shape (bottom right).}
    \label{fig:barrier2m}
\end{figure}

Figure~\ref{fig:barrier2m} shows a different view of these correlations.  At fixed mass, protohalos with $\epsilon > \sqrt{\epsilon_c^2 + v_+^2}$ tend to have larger $z_{50}$.  The bottom panels show that there is also a correlation with the concentration of the evolved halo (bottom left) but less so for its shape (bottom right).

At fixed mass, protohalos with the largest $p_{\cal E}$ become the roundest halos. Such protohalos have one large axis and two others that are similar (i.e., their energy tensors are prolate rather than oblate), making $p_{\cal E}\approx e_{\cal E}$, and $q_{\cal E}^2\approx (2e_{\cal E})^2$.  The protohalos of the roundest halos also tend to have the largest $e_{\cal E}$ and $q_{\cal E}^2$, suggesting that the roundest objects result from non-spherical initial patches (magenta bands in $e_{\cal M}$ and $q^2_{\cal M}$ in the bottom panels of Figure~\ref{fig:tensorM}) that were squeezed equally in two directions, and even more in the third.  

\begin{figure}
    \centering
    \includegraphics[width=0.45\linewidth]{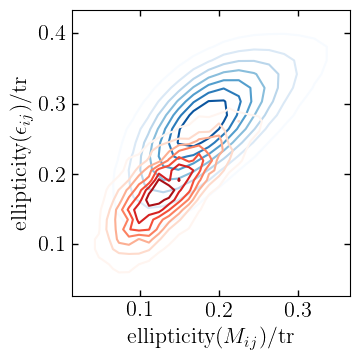}
    \includegraphics[width=0.45\linewidth]{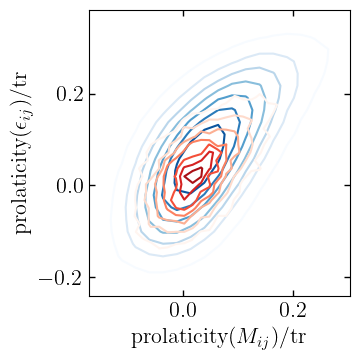}
    \caption{At fixed mass (colors same as previous figure) the ellipticities (left) and prolatenesses (right) of the mass and energy tensors are correlated, indicating that the shapes of the mass and energy tensors are similar.}
    \label{fig:ells}
\end{figure}

Ref.\cite{eshape23} argues that the shapes of the mass and energy tensors should be qualitatively similar:  if the energy tensor is prolate, then the mass tensor should be as well.  Figure~\ref{fig:ells} shows that, indeed, the ellipticities and prolatenesses are quite well correlated.  
However, while larger $e_\epsilon$ tend to have larger $z_{50}$ (top panels of Figure~\ref{fig:tensorE}), larger $e_{\cal{M}}$ have smaller $z_{50}$ (top panels of Figure~\ref{fig:tensorM}).  This latter correlation for the protohalo shape is qualitatively similar to that of the evolved halos: rounder shapes are associated with earlier assembly.  Therefore, if we were to write 
$z_{50}\approx a_{z_{50}} e_\epsilon + b_{z_{50}} e_{\cal M}$, 
then $a_{z_{50}}\ge 0$ whereas $b_{z_{50}}\le 0$. This makes good physical sense, since the corresponding eigenvectors are very well aligned (left-hand panel of Figure~3 of \cite{mds2024}) so anisotropic squeezing leads to earlier assembly, unless the object was already elliptical.  Note that this is also consistent with the tendency for $\delta$ or $\epsilon$ to be larger if the anisotropy of ${\cal D}$ or ${\cal E}$ is large \citep{smt2001, rotinv}, as large values of $\delta$ are associated with earlier assembly (Figures~\ref{fig:tensorD} and~\ref{fig:tensorE}).
The concentration is only slightly different: larger $e_{\cal M}$ are less concentrated (qualitatively like $z_{50}$), whereas $e_{\cal E}$ and concentration are almost uncorrelated.  Despite this difference, if we write 
  ${\rm conc}\approx a_{\rm conc} e_{\cal E} + b_{\rm conc} e_{\cal M}$, 
then $a_{\rm conc}\ge 0$ whereas $b_{\rm conc}\le 0$. 

\section{Conclusions}
\label{sec:conclusions}

We introduced a new measure of halo assembly and demonstrated its similarity to the commonly used $z_{50}$ (Figures~\ref{fig:MAH} and\ref{fig:W1-z50}). This new measure, derived from an integral of the mass accretion history (Eqn.~\ref{eq:MAH}), is expected to be slightly more robust.

We then used the rotational invariants (Eqs.~\ref{eq:invariants} and \ref{eq:epv}) of the mass, deformation, and energy shear tensors (Eqs.~\ref{eq:Mij}, \ref{eq:Dij}, and~\ref{eq:Eij}) to explore the correlation between protohalo structure, halo formation, and halo shape (Figures~\ref{fig:tensorM}--\ref{fig:tensorE}). Our key findings are:

\concsection{Correlation of shapes} The mass and energy shear tensors have similar shapes:  if one is more elliptical, then the other is as well; if one is prolate, then so is the other (Figure~\ref{fig:ells}).  This agrees qualitatively with the variational principle for protohalo shapes described in \cite{eshape23}.

\concsection{Shape Evolution} The roundest halos today did not originate from the roundest protohalo patches, but had initial ellipticities $e_{\cal M} \sim 0.2$ (bottom row of Figure~\ref{fig:tensorM}).

\concsection{Formation Times} At lower masses, the objects that end up being roundest typically have median formation times.

\concsection{Eigenvalue Patterns} Objects with $p_{\cal M}\sim 0$ tend to form earliest:  prolate or oblate objects (i.e. whose mass tensors have two equal eigenvalues) form later.

\concsection{Tensor Trends} For the deformation and energy shear tensors, larger trace, shear, and ellipticity correlate with earlier formation, higher concentration, and rounder shapes compared to counterparts of the same mass (Figures~\ref{fig:tensorD} and\ref{fig:tensorE}), whereas the mass tensor displays opposite trends.

\concsection{Invariant Correlations} While the third invariant $U_3/q^3$ shows little correlation with halo properties at higher masses, lower mass objects with larger values of this invariant form later, are less concentrated, but are rounder (middle column of Figures~\ref{fig:tensorD} and \ref{fig:tensorE}, and Figure~\ref{fig:3rd}); these trends are stronger and similar to those observed for the mass tensor.

\concsection{Energy Peaks Model} Within the energy peaks model of halo formation, objects above the $\epsilon - v_+$ relation (Eqn.~\ref{eq:epsv+}) form earlier and exhibit higher concentrations (Figures~\ref{fig:barrierczy} and \ref{fig:barrier2m}), with the final shape correlating with $v_+$ only for the most massive objects (Figure~\ref{fig:barrierczy}).

The observed correlation between the final halo shape (and to a lesser extent, assembly time) and the third invariant $\mu_3\equiv (9/2)(U/q)^3$ warrants further discussion. Previous studies have shown that correlations between protohalo properties and the first and second invariants (the trace and the amplitude of the traceless shear) can be efficiently estimated using orthogonal polynomials: Hermite and Laguerre polynomials for a Gaussian field, respectively \cite{mps12, vd13, cps17}. Given that $\mu_3$ is bounded between $[-1,1]$, Legendre polynomials are the appropriate basis functions for estimating the bias associated with the third invariant \cite{lmd16}. Hence, our work motivates the application of Legendre polynomial weightings of the large scale $\mu_3$ field, centered on protohalo positions, for bias factor estimation.  More generally, our work contributes to ongoing discussions of the physical origin of halo assembly or secondary bias \cite{st04,fw10,shapeSigma,Blazek_shapebias2015,Lee_shapebias2024,Montero-Dorta2025}.

Finally, Optimal Transport reconstructions have shown promise in recreating the initial shapes of protohalos \citep{PRLhalos, OTrsd, OTmt}. To test these reconstructions quantitatively, it will be interesting to compare the OT-reconstructed shapes to the corresponding protohalo mass tensors $\cal{M}$ and examine the displacements which give rise to shape evolution in comparison to $\cal{D}$ and $\cal{E}$. Accurate reconstructions can provide direct estimates of the formation history and shape evolution of each halo. This will also inform the impact of shape and assembly bias on the cosmological information provided by these objects, an area of ongoing work.

\section*{Acknowledgements}

We thank Yan-Chuan Cai for valuable discussions and comments, and Sownak Bose for guidance in accessing the AbacusSummit simulations. FN gratefully acknowledges support from the Yale Center for Astronomy and Astrophysics Prize Postdoctoral Fellowship. RKS is grateful to the ICTP for its hospitality in 2024.

\section*{Data Availability}
The simulation data and post-processed quantities used in this work can be shared on reasonable request to the authors.




\bibliographystyle{mnras}
\bibliography{mybib} 

@ARTICLE{mds2024,
       author = {{Musso}, Marcello and {Despali}, Giulia and {Sheth}, Ravi K.},
        title = "{The energy shear of protohaloes}",
      journal = {\aap},
         year = 2024,
        month = oct,
       volume = {690},
          eid = {A214},
        pages = {A214},
          doi = {10.1051/0004-6361/202450985},
archivePrefix = {arXiv},
       eprint = {2405.20207},
 primaryClass = {astro-ph.CO},
       adsurl = {https://ui.adsabs.harvard.edu/abs/2024A&A...690A.214M}
}

@ARTICLE{rotinv,
       author = {{Musso}, Marcello and {Sheth}, Ravi K.},
        title = "{The role of energy shear in the collapse of protohaloes}",
      journal = {arXiv e-prints},
         year = 2024,
        month = oct,
          eid = {arXiv:2410.06289},
        pages = {arXiv:2410.06289},
          doi = {10.48550/arXiv.2410.06289},
archivePrefix = {arXiv},
       eprint = {2410.06289},
 primaryClass = {astro-ph.CO},
       adsurl = {https://ui.adsabs.harvard.edu/abs/2024arXiv241006289M}
}

@ARTICLE{Garrison2021,
       author = {{Garrison}, Lehman H. and {Eisenstein}, Daniel J. and {Ferrer}, Douglas and {Maksimova}, Nina A. and {Pinto}, Philip A.},
        title = "{The ABACUS cosmological N-body code}",
      journal = {\mnras},
         year = 2021,
        month = nov,
       volume = {508},
       number = {1},
        pages = {575-596},
          doi = {10.1093/mnras/stab2482},
archivePrefix = {arXiv},
       eprint = {2110.11392},
 primaryClass = {astro-ph.CO},
       adsurl = {https://ui.adsabs.harvard.edu/abs/2021MNRAS.508..575G}
}

@ARTICLE{Maksimova2021,
       author = {{Maksimova}, Nina A. and {Garrison}, Lehman H. and {Eisenstein}, Daniel J. and {Hadzhiyska}, Boryana and {Bose}, Sownak and {Satterthwaite}, Thomas P.},
        title = "{ABACUSSUMMIT: a massive set of high-accuracy, high-resolution N-body simulations}",
      journal = {\mnras},
         year = 2021,
        month = dec,
       volume = {508},
       number = {3},
        pages = {4017-4037},
          doi = {10.1093/mnras/stab2484},
archivePrefix = {arXiv},
       eprint = {2110.11398},
 primaryClass = {astro-ph.CO},
       adsurl = {https://ui.adsabs.harvard.edu/abs/2021MNRAS.508.4017M}
}

@ARTICLE{Hadzhiyska2022,
       author = {{Hadzhiyska}, Boryana and {Eisenstein}, Daniel and {Bose}, Sownak and {Garrison}, Lehman H. and {Maksimova}, Nina},
        title = "{COMPASO: A new halo finder for competitive assignment to spherical overdensities}",
      journal = {\mnras},
         year = 2022,
        month = jan,
       volume = {509},
       number = {1},
        pages = {501-521},
          doi = {10.1093/mnras/stab2980},
archivePrefix = {arXiv},
       eprint = {2110.11408},
 primaryClass = {astro-ph.CO},
       adsurl = {https://ui.adsabs.harvard.edu/abs/2022MNRAS.509..501H}
}

@ARTICLE{Bose2022,
       author = {{Bose}, Sownak and {Eisenstein}, Daniel J. and {Hadzhiyska}, Boryana and {Garrison}, Lehman H. and {Yuan}, Sihan},
        title = "{Constructing high-fidelity halo merger trees in ABACUSSUMMIT}",
      journal = {\mnras},
         year = 2022,
        month = may,
       volume = {512},
       number = {1},
        pages = {837-854},
          doi = {10.1093/mnras/stac555},
archivePrefix = {arXiv},
       eprint = {2110.11409},
 primaryClass = {astro-ph.CO},
       adsurl = {https://ui.adsabs.harvard.edu/abs/2022MNRAS.512..837B}
}

@ARTICLE{mps12,
       author = {{Musso}, Marcello and {Paranjape}, Aseem and {Sheth}, Ravi K.},
        title = "{Scale-dependent halo bias in the excursion set approach}",
      journal = {\mnras},
         year = 2012,
        month = dec,
       volume = {427},
       number = {4},
        pages = {3145-3158},
          doi = {10.1111/j.1365-2966.2012.21903.x},
archivePrefix = {arXiv},
       eprint = {1205.3401},
 primaryClass = {astro-ph.CO},
       adsurl = {https://ui.adsabs.harvard.edu/abs/2012MNRAS.427.3145M}
}

@ARTICLE{cps17,
       author = {{Castorina}, Emanuele and {Paranjape}, Aseem and {Sheth}, Ravi K.},
        title = "{Constraints on halo formation from cross-correlations with correlated variables}",
      journal = {\mnras},
         year = 2017,
        month = jul,
       volume = {468},
       number = {4},
        pages = {3813-3827},
          doi = {10.1093/mnras/stx701},
archivePrefix = {arXiv},
       eprint = {1611.03613},
 primaryClass = {astro-ph.CO},
       adsurl = {https://ui.adsabs.harvard.edu/abs/2017MNRAS.468.3813C}
}

@ARTICLE{vd13,
       author = {{Desjacques}, Vincent},
        title = "{Local bias approach to the clustering of discrete density peaks}",
      journal = {\prd},
         year = 2013,
        month = feb,
       volume = {87},
       number = {4},
          eid = {043505},
        pages = {043505},
          doi = {10.1103/PhysRevD.87.043505},
archivePrefix = {arXiv},
       eprint = {1211.4128},
 primaryClass = {astro-ph.CO},
       adsurl = {https://ui.adsabs.harvard.edu/abs/2013PhRvD..87d3505D}
}

@ARTICLE{lmd16,
       author = {{Lazeyras}, Titouan and {Musso}, Marcello and {Desjacques}, Vincent},
        title = "{Lagrangian bias of generic large-scale structure tracers}",
      journal = {\prd},
         year = 2016,
        month = mar,
       volume = {93},
       number = {6},
          eid = {063007},
        pages = {063007},
          doi = {10.1103/PhysRevD.93.063007},
archivePrefix = {arXiv},
       eprint = {1512.05283},
 primaryClass = {astro-ph.CO},
       adsurl = {https://ui.adsabs.harvard.edu/abs/2016PhRvD..93f3007L}
}

@ARTICLE{Allen2011,
       author = {{Allen}, Steven W. and {Evrard}, August E. and {Mantz}, Adam B.},
        title = "{Cosmological Parameters from Observations of Galaxy Clusters}",
      journal = {\araa},
         year = 2011,
        month = sep,
       volume = {49},
       number = {1},
        pages = {409-470},
          doi = {10.1146/annurev-astro-081710-102514},
archivePrefix = {arXiv},
       eprint = {1103.4829},
 primaryClass = {astro-ph.CO},
       adsurl = {https://ui.adsabs.harvard.edu/abs/2011ARA&A..49..409A}
}

@ARTICLE{Bocquet2024,
       author = {{Bocquet}, S. and {Grandis}, S. and {Bleem}, L.~E. and {Klein}, M. and {Mohr}, J.~J. and {Schrabback}, T. and {Abbott}, T.~M.~C. and {Ade}, P.~A.~R. and {Aguena}, M. and {Alarcon}, A. and {Allam}, S. and {Allen}, S.~W. and {Alves}, O. and {Amon}, A. and {Anderson}, A.~J. and {Annis}, J. and {Ansarinejad}, B. and {Austermann}, J.~E. and {Avila}, S. and {Bacon}, D. and {Bayliss}, M. and {Beall}, J.~A. and {Bechtol}, K. and {Becker}, M.~R. and {Bender}, A.~N. and {Benson}, B.~A. and {Bernstein}, G.~M. and {Bhargava}, S. and {Bianchini}, F. and {Brodwin}, M. and {Brooks}, D. and {Bryant}, L. and {Campos}, A. and {Canning}, R.~E.~A. and {Carlstrom}, J.~E. and {Carnero Rosell}, A. and {Carrasco Kind}, M. and {Carretero}, J. and {Castander}, F.~J. and {Cawthon}, R. and {Chang}, C.~L. and {Chang}, C. and {Chaubal}, P. and {Chen}, R. and {Chiang}, H.~C. and {Choi}, A. and {Chou}, T. -L. and {Citron}, R. and {Corbett Moran}, C. and {Cordero}, J. and {Costanzi}, M. and {Crawford}, T.~M. and {Crites}, A.~T. and {da Costa}, L.~N. and {Pereira}, M.~E.~S. and {Davis}, C. and {Davis}, T.~M. and {DeRose}, J. and {Desai}, S. and {de Haan}, T. and {Diehl}, H.~T. and {Dobbs}, M.~A. and {Dodelson}, S. and {Doux}, C. and {Drlica-Wagner}, A. and {Eckert}, K. and {Elvin-Poole}, J. and {Everett}, S. and {Everett}, W. and {Ferrero}, I. and {Fert{\'e}}, A. and {Flores}, A.~M. and {Frieman}, J. and {Gallicchio}, J. and {Garc{\'\i}a-Bellido}, J. and {Gatti}, M. and {George}, E.~M. and {Giannini}, G. and {Gladders}, M.~D. and {Gruen}, D. and {Gruendl}, R.~A. and {Gupta}, N. and {Gutierrez}, G. and {Halverson}, N.~W. and {Harrison}, I. and {Hartley}, W.~G. and {Herner}, K. and {Hinton}, S.~R. and {Holder}, G.~P. and {Hollowood}, D.~L. and {Holzapfel}, W.~L. and {Honscheid}, K. and {Hrubes}, J.~D. and {Huang}, N. and {Hubmayr}, J. and {Huff}, E.~M. and {Huterer}, D. and {Irwin}, K.~D. and {James}, D.~J. and {Jarvis}, M. and {Khullar}, G. and {Kim}, K. and {Knox}, L. and {Kraft}, R. and {Krause}, E. and {Kuehn}, K. and {Kuropatkin}, N. and {K{\'e}ruzor{\'e}}, F. and {Lahav}, O. and {Lee}, A.~T. and {Leget}, P. -F. and {Li}, D. and {Lin}, H. and {Lowitz}, A. and {MacCrann}, N. and {Mahler}, G. and {Mantz}, A. and {Marshall}, J.~L. and {McCullough}, J. and {McDonald}, M. and {McMahon}, J.~J. and {Mena-Fern{\'a}ndez}, J. and {Menanteau}, F. and {Meyer}, S.~S. and {Miquel}, R. and {Montgomery}, J. and {Myles}, J. and {Natoli}, T. and {Navarro-Alsina}, A. and {Nibarger}, J.~P. and {Noble}, G.~I. and {Novosad}, V. and {Ogando}, R.~L.~C. and {Omori}, Y. and {Padin}, S. and {Pandey}, S. and {Paschos}, P. and {Patil}, S. and {Pieres}, A. and {Plazas Malag{\'o}n}, A.~A. and {Porredon}, A. and {Prat}, J. and {Pryke}, C. and {Raveri}, M. and {Reichardt}, C.~L. and {Roberson}, J. and {Rollins}, R.~P. and {Romero}, C. and {Roodman}, A. and {Ruhl}, J.~E. and {Rykoff}, E.~S. and {Saliwanchik}, B.~R. and {Salvati}, L. and {S{\'a}nchez}, C. and {Sanchez}, E. and {Sanchez Cid}, D. and {Saro}, A. and {Schaffer}, K.~K. and {Secco}, L.~F. and {Sevilla-Noarbe}, I. and {Sharon}, K. and {Sheldon}, E. and {Shin}, T. and {Sievers}, C. and {Smecher}, G. and {Smith}, M. and {Somboonpanyakul}, T. and {Sommer}, M. and {Stalder}, B. and {Stark}, A.~A. and {Stephen}, J. and {Strazzullo}, V. and {Suchyta}, E. and {Tarle}, G. and {To}, C. and {Troxel}, M.~A. and {Tucker}, C. and {Tutusaus}, I. and {Varga}, T.~N. and {Veach}, T. and {Vieira}, J.~D. and {Vikhlinin}, A. and {von der Linden}, A. and {Wang}, G. and {Weaverdyck}, N. and {Weller}, J. and {Whitehorn}, N. and {Wu}, W.~L.~K. and {Yanny}, B. and {Yefremenko}, V. and {Yin}, B. and {Young}, M. and {Zebrowski}, J.~A. and {Zhang}, Y. and {Zohren}, H. and {Zuntz}, J. and {(SPT} and {DES Collaborations)}},
        title = "{SPT clusters with DES and HST weak lensing. II. Cosmological constraints from the abundance of massive halos}",
      journal = {\prd},
         year = 2024,
        month = oct,
       volume = {110},
       number = {8},
          eid = {083510},
        pages = {083510},
          doi = {10.1103/PhysRevD.110.083510},
archivePrefix = {arXiv},
       eprint = {2401.02075},
 primaryClass = {astro-ph.CO},
       adsurl = {https://ui.adsabs.harvard.edu/abs/2024PhRvD.110h3510B}
}

@ARTICLE{OTmt,
       author = {{Nikakhtar}, Farnik and {Sheth}, Ravi K. and {Padmanabhan}, Nikhil and {L{\'e}vy}, Bruno and {Mohayaee}, Roya},
        title = "{Displacement field analysis via optimal transport: Multitracer approach to cosmological reconstruction}",
      journal = {\prd},
         year = 2024,
        month = jun,
       volume = {109},
       number = {12},
          eid = {123512},
        pages = {123512},
          doi = {10.1103/PhysRevD.109.123512},
archivePrefix = {arXiv},
       eprint = {2403.11951},
 primaryClass = {astro-ph.CO},
       adsurl = {https://ui.adsabs.harvard.edu/abs/2024PhRvD.109l3512N}
}

@ARTICLE{OTrsd,
       author = {{Nikakhtar}, Farnik and {Padmanabhan}, Nikhil and {L{\'e}vy}, Bruno and {Sheth}, Ravi K. and {Mohayaee}, Roya},
        title = "{Optimal transport reconstruction of biased tracers in redshift space}",
      journal = {\prd},
         year = 2023,
        month = oct,
       volume = {108},
       number = {8},
          eid = {083534},
        pages = {083534},
          doi = {10.1103/PhysRevD.108.083534},
archivePrefix = {arXiv},
       eprint = {2307.03671},
 primaryClass = {astro-ph.CO},
       adsurl = {https://ui.adsabs.harvard.edu/abs/2023PhRvD.108h3534N}
}

@ARTICLE{PRLhalos,
       author = {{Nikakhtar}, Farnik and {Sheth}, Ravi K. and {L{\'e}vy}, Bruno and {Mohayaee}, Roya},
        title = "{Optimal Transport Reconstruction of Baryon Acoustic Oscillations}",
      journal = {\prl},
         year = 2022,
        month = dec,
       volume = {129},
       number = {25},
          eid = {251101},
        pages = {251101},
          doi = {10.1103/PhysRevLett.129.251101},
archivePrefix = {arXiv},
       eprint = {2203.01868},
 primaryClass = {astro-ph.CO},
       adsurl = {https://ui.adsabs.harvard.edu/abs/2022PhRvL.129y1101N}
}

@article{Blazek_shapebias2015,
    author = "Blazek, Jonathan and Vlah, Zvonimir and Seljak, Uro\v{s}",
    title = "{Tidal alignment of galaxies}",
    eprint = "1504.02510",
    archivePrefix = "arXiv",
    primaryClass = "astro-ph.CO",
    doi = "10.1088/1475-7516/2015/08/015",
    journal = "JCAP",
    volume = "08",
    pages = "015",
    year = "2015"
}

@ARTICLE{Lee_shapebias2024,
       author = {{Lee}, Jounghun and {Moon}, Jun-Sung},
        title = "{The dependence of halo bias on the protohalo shape alignment with the initial tidal field}",
      journal = {\jcap},
         year = 2024,
        month = oct,
       volume = {2024},
       number = {10},
          eid = {102},
        pages = {102},
          doi = {10.1088/1475-7516/2024/10/102},
archivePrefix = {arXiv},
       eprint = {2406.11182},
 primaryClass = {astro-ph.CO},
       adsurl = {https://ui.adsabs.harvard.edu/abs/2024JCAP...10..102L}
}

@ARTICLE{shapeSigma,
       author = {{Ragone-Figueroa}, C. and {Plionis}, M. and {Merch{\'a}n}, M. and {Gottl{\"o}ber}, S. and {Yepes}, G.},
        title = "{The relation between halo shape, velocity dispersion and formation time}",
      journal = {\mnras},
         year = 2010,
        month = sep,
       volume = {407},
       number = {1},
        pages = {581-589},
          doi = {10.1111/j.1365-2966.2010.16935.x},
archivePrefix = {arXiv},
       eprint = {1005.1870},
 primaryClass = {astro-ph.CO},
       adsurl = {https://ui.adsabs.harvard.edu/abs/2010MNRAS.407..581R}
}

@article{dts2013,
    author = "Despali, Giulia and Tormen, Giuseppe and Sheth, Ravi K.",
    title = "{Ellipsoidal halo finders and implications for models of triaxial halo formation}",
    eprint = "1212.4157",
    archivePrefix = "arXiv",
    primaryClass = "astro-ph.CO",
    doi = "10.1093/mnras/stt235",
    journal = {\mnras},
    volume = "431",
    number = "2",
    pages = "1143--1159",
    year = "2013"
}

@ARTICLE{doroshkevich1970,
       author = {{Doroshkevich}, A.~G.},
        title = "{Spatial structure of perturbations and origin of galactic rotation in fluctuation theory}",
      journal = {Astrophysics},
         year = 1970,
        month = oct,
       volume = {6},
       number = {4},
        pages = {320-330},
          doi = {10.1007/BF01001625},
       adsurl = {https://ui.adsabs.harvard.edu/abs/1970Ap......6..320D}
}

@ARTICLE{nfw1997,
       author = {{Navarro}, Julio F. and {Frenk}, Carlos S. and {White}, Simon D.~M.},
        title = "{A Universal Density Profile from Hierarchical Clustering}",
      journal = {\apj},
         year = 1997,
        month = dec,
       volume = {490},
       number = {2},
        pages = {493-508},
          doi = {10.1086/304888},
archivePrefix = {arXiv},
       eprint = {astro-ph/9611107},
 primaryClass = {astro-ph},
       adsurl = {https://ui.adsabs.harvard.edu/abs/1997ApJ...490..493N}
}

@article{eshape23,
    author = "Musso, Marcello and Sheth, Ravi K.",
    title = "{Getting in shape with minimal energy: a variational principle for protohaloes}",
    eprint = "2303.02142",
    archivePrefix = "arXiv",
    primaryClass = "astro-ph.GA",
    doi = "10.1093/mnrasl/slad044",
    journal = "\mnras",
    volume = "523",
    number = "1",
    pages = "L4--L8",
    year = "2023"
}

@article{epeaks,
    author = "Musso, Marcello and Sheth, Ravi K.",
    title = "{Excursion set peaks in energy as a model for haloes}",
    eprint = "1907.09147",
    archivePrefix = "arXiv",
    primaryClass = "astro-ph.CO",
    doi = "10.1093/mnras/stab2640",
    journal = {\mnras},
    volume = "508",
    number = "3",
    pages = "3634--3648",
    year = "2021"
}

@ARTICLE{borzy_colltime14,
       author = {{Borzyszkowski}, Mikolaj and {Ludlow}, Aaron D. and {Porciani}, Cristiano},
        title = "{The formation of cold dark matter haloes - II. Collapse time and tides}",
      journal = {\mnras},
         year = 2014,
        month = dec,
       volume = {445},
       number = {4},
        pages = {4124-4136},
          doi = {10.1093/mnras/stu2033},
archivePrefix = {arXiv},
       eprint = {1405.7367},
 primaryClass = {astro-ph.CO},
       adsurl = {https://ui.adsabs.harvard.edu/abs/2014MNRAS.445.4124B}
}

@Article{bbks86,
     author    = "Bardeen, James M. and Bond, J. R. and Kaiser, Nick and
                  Szalay, A. S.",
     title     = "{The Statistics of Peaks of Gaussian Random Fields}",
     journal   = "\apj",
     volume    = "304",
     year      = "1986",
     pages     = "15-61",
     doi       = "10.1086/164143",
     SLACcitation  = "%%CITATION = ASJOA,304,15;%%"
}

@ARTICLE{bm96,
   author = {{Bond}, J.~R. and {Myers}, S.~T.},
    title = "{The Peak-Patch Picture of Cosmic Catalogs. I. Algorithms}",
  journal = {\apjs},
     year = 1996,
    month = mar,
   volume = 103,
    pages = {1},
      doi = {10.1086/192267},
   adsurl = {http://adsabs.harvard.edu/abs/1996ApJS..103....1B}
}

@ARTICLE{fw10,
       author = {{Faltenbacher}, Andreas and {White}, Simon D.~M.},
        title = "{Assembly Bias and the Dynamical Structure of Dark Matter Halos}",
      journal = {\apj},
         year = 2010,
        month = jan,
       volume = {708},
       number = {1},
        pages = {469-473},
          doi = {10.1088/0004-637X/708/1/469},
archivePrefix = {arXiv},
       eprint = {0909.4302},
 primaryClass = {astro-ph.CO},
       adsurl = {https://ui.adsabs.harvard.edu/abs/2010ApJ...708..469F}
}

@ARTICLE{Montero-Dorta2025,
       author = {{Montero-Dorta}, Antonio D. and {Contreras}, Sergio and {Artale}, M. Celeste and {Rodriguez}, Facundo and {Favole}, Ginevra},
        title = "{Exploring the physical origins of halo assembly bias from early times}",
      journal = {\aap},
         year = 2025,
        month = mar,
       volume = {695},
          eid = {A159},
        pages = {A159},
          doi = {10.1051/0004-6361/202452709},
archivePrefix = {arXiv},
       eprint = {2410.18319},
 primaryClass = {astro-ph.CO},
       adsurl = {https://ui.adsabs.harvard.edu/abs/2025A&A...695A.159M}
}

@ARTICLE{psd13,
   author = {{Paranjape}, A. and {Sheth}, R.~K. and {Desjacques}, V.},
    title = "{Excursion set peaks: a self-consistent model of dark halo abundances and clustering}",
  journal = {\mnras},
archivePrefix = "arXiv",
   eprint = {1210.1483},
     year = 2013,
    month = may,
   volume = 431,
    pages = {1503-1512},
      doi = {10.1093/mnras/stt267},
   adsurl = {http://adsabs.harvard.edu/abs/2013MNRAS.431.1503P}
}

@Article{smt2001,
     author    = "Sheth, Ravi K. and Mo, Houjon and Tormen, Giuseppe",
     title     = "{Ellipsoidal collapse and an improved model for the number and spatial distribution of dark matter haloes}",
     journal   = "MNRAS",
     volume    = "323",
     year      = "2001",
     pages     = "1",
     doi       = "10.1046/j.1365-8711.2001.04006.x",
     eprint    = "9907024",
     archivePrefix = "arXiv",
     primaryClass  =  "astro-ph.CO",
     SLACcitation  = "%%CITATION = 9907024;%%"
}

@ARTICLE{st04,
   author = {{Sheth}, R.~K. and {Tormen}, G.},
    title = "{On the environmental dependence of halo formation}",
  journal = {\mnras},
   eprint = {astro-ph/0402237},
     year = 2004,
    month = jun,
   volume = 350,
    pages = {1385-1390},
      doi = {10.1111/j.1365-2966.2004.07733.x},
   adsurl = {http://adsabs.harvard.edu/abs/2004MNRAS.350.1385S}
}



\appendix

\section{Interpretation of \texorpdfstring{$A$}{A} as a weighted integral of the MAH}

The main text introduced our new measure, $A$, of the mass assembly history.  It was inspired by optimal transport discussions of the difference between two distributions.
Let the cumulative distributions be $F_1(x)$ and $F_2(x)$.  Then the Wasserstein-$p$ distance, with $p=1$ is 
\begin{equation}
    W_1 \equiv \int dx\,|F_1(\le x) - F_2(\le x)|.
\end{equation}
In our case, the `distributions' are $dm(z)/dz$, where all masses are written in units of the final mass, and we are interested in quantifying how different the MAH distribution of a given halo is from the median MAH of all halos of the same final mass.  Since $m(\ge z)$ is like a cumulative distribution which would integrate to 1 at $z=z_{\rm final}$ (=0 for the halos in the main text), our $A$ is like $W_1$, but without the absolute value sign.  For our problem, 
\begin{align}
    A_z &= \int_{z_{\rm init}}^{z_{\rm fin}} \frac{dz}{z_{\rm fin} - z_{\rm init}}
         \int_{z_{\rm init}}^z dz'\,\left(\frac{dm_1}{dz'} - \frac{dm_2}{dz'}\right)
         \nonumber\\
      &= \int_{z_{\rm init}}^{z_{\rm fin}} dz'\,\left(\frac{dm_1}{dz'} - \frac{dm_2}{dz'}\right)
         \frac{z_{\rm fin} - z'}{z_{\rm fin} - z_{\rm init}}
\end{align}
With $z_{\rm fin}=0$, this clearly just weights the mass accretion by $z$: earlier accretion (higher $z$) contributes more.  

Had we decided to work with $a/a_{\rm fin}=(1+z_{\rm fin})/(1+z)$ instead, we would have had 
\begin{align}
    A_a &= \int_{a_{\rm init}}^1 \frac{da}{1-a_{\rm init}}
         \int_{a_{\rm init}}^a da'\,\left(\frac{dm_1}{da'} - \frac{dm_2}{da'}\right)
         \nonumber\\
      &= \int_{a_{\rm init}}^1 da'\,\left(\frac{dm_1}{da'} - \frac{dm_2}{da'}\right)
         \frac{1 - a'}{1 - a_{\rm init}}
\end{align}
This also weights early times (small $a'$) more than late times, but not as strongly as when working with $z$.  

To see that this matters, note that the main text used 
\begin{equation}
 m(z) = \int_{z_{\rm init}}^z dz\,\frac{dm}{dz} = \int_{a_{\rm init}}^a da\,\frac{dm}{da} = m(a).
\end{equation}
Since
\begin{equation}
     \int dz\,m(z) = \int da\, \frac{dz}{da}\,m(a)\ne \int da\,m(a), 
\end{equation}
we will have $A_z\ne A_a$.


\bsp	
\label{lastpage}
\end{document}